\documentclass[preprint,superscriptaddress,aps]{revtex4}
\usepackage{amssymb}
\usepackage{amsmath,amssymb,graphicx}
\usepackage{graphicx}
\usepackage{dcolumn}
\usepackage{bm}
\def\be{\begin{equation}}
\def\ee{\end{equation}}
\def\bea{\begin{eqnarray}}
\def\eea{\end{eqnarray}}

\def\lsim{\raise0.3ex\hbox{$\;<$\kern-0.75em\raise-1.1ex\hbox{$\sim\;$}}}
\def\gsim{\raise0.3ex\hbox{$\;>$\kern-0.75em\raise-1.1ex\hbox{$\sim\;$}}}

\def\ie{{\it i.e.}}

\usepackage{pstricks}
\begin{document}
\title{{\fontsize{13}{13}\selectfont{MSSM Dark Matter in Light of Higgs and LUX Results}}}
\author{W. Abdallah}
\affiliation{{\fontsize{10}{10}\selectfont{Center for Fundamental Physics, Zewail City of Science and Technology, 6 October City, Giza 12588, Egypt.}}}
\affiliation{{\fontsize{10}{10}\selectfont{Department of Mathematics, Faculty of Science, Cairo University, Giza 12613, Egypt.}}}
\author{S. Khalil}
\affiliation{{\fontsize{10}{10}\selectfont{Center for Fundamental Physics, Zewail City of Science and Technology, 6 October City, Giza 12588, Egypt.}}}

\begin{center}
\begin{abstract}

The constraints imposed on the Minimal Supersymmetric Standard Model (MSSM) parameter space by the Large Hadron Collider (LHC) Higgs mass limit and gluino mass lower bound are revisited. We also analyze the thermal relic abundance of lightest neutralino, which is the Lightest Supersymmetric Particle (LSP). We show that the combined LHC and relic abundance constraints rule out most of the MSSM parameter space except a very narrow region with very large $\tan \beta~(\sim 50)$.  Within this region, we emphasize that the spin-independent scattering cross section of the LSP with a proton is less than the latest Large Underground Xenon (LUX) limit by at least two order of magnitudes. Finally, we argue that non-thermal Dark Matter (DM) scenario may relax the constraints imposed on the MSSM parameter space. Namely, the following regions are obtained: $m_0\simeq {\cal O}(4)$~TeV and $m_{1/2}\simeq 600$~GeV for low $\tan\beta~(\sim 10)$; $m_0\sim m_{1/2} \simeq {\cal O}(1)$~TeV or $m_0 \simeq {\cal O}(4)$~TeV and $m_{1/2} \simeq 700$~GeV for large $\tan \beta~(\sim 50)$.

\end{abstract}
\end{center}
\maketitle
\section{Introduction}

The most recent observations by the Planck satellite confirmed that $26.8 \%$ of the universe content in the form of DM and the usual visible matter only accounts for $5\%$ \cite{Ade:2013zuv}. The LSP remains one of the best candidates for the DM \cite{Jungman:1995df,Bertone:2004pz}. It is a Weakly Interacting Massive Particle (WIMP) that can naturally account for the observed relic density of DM.

Despite the absence of direct experimental verification, Supersymmetry (SUSY) is still the most promising candidate for a unified theory beyond the Standard Model (SM).
SUSY is a generalization of the space-time symmetries of the quantum field theory that links the matter particles (quarks and leptons) with the force-carrying particles, and implies that there are additional `superparticles' necessary to complete the symmetry. In this regards, SUSY solves the problem of the quadratic divergence in the Higgs sector of the SM in a very elegant natural way. The most simple supersymmetric extension of the SM, which is the most widely studied, is know as the MSSM \cite{susyreviews1,susyreviews2,susybooks}. In this model, certain universality of soft SUSY breaking terms is assumed at grand unification scale. Therefore, the SUSY spectrum is determined by the following four parameters: universal scalar mass $m_0$, universal gaugino mass $m_{1/2}$, universal trilinear coupling $A_0$,  the ratio of the vacuum expectation values of Higgs bosons $\tan \beta$. 
In addition, due to $R$-parity conservation, SUSY particles are produced or destroyed only in pairs and therefore the LSP is absolutely stable, implying that it might constitute a possible candidate for DM, as first suggested by Goldberg in 1983 \cite{Goldberg:1983nd}. So although the original motivation of SUSY has nothing to do with the DM problem, it turns out that it provides a stable neutral particle and, hence, a candidate for solving the DM problem.

The landmark discovery of the SM-like Higgs boson at the LHC, with mass $\sim 125$~GeV \cite{Aad:2012tfa}, might be an indication for the presence of SUSY. Indeed, the MSSM predicts that there is an upper bound of 130~GeV on the Higgs mass. However, this mass of lightest Higgs boson implies that the SUSY particles are quite heavy.  This may justify the negative searches for SUSY at the LHC-run I \cite{SUSYRUN1}. However, it is clearly generating  a new `little hierarchy problem'.

Moreover, the relic density data \cite{Ade:2013zuv} and upper limits on the DM scattering cross sections on nuclei (LUX \cite{Akerib:2013tjd} and other direct detection experiments \cite{Aprile:2012zx}) impose stringent constraints on the parameter space of the MSSM \cite{dd_msugra}. In fact, combining the collider, astrophysics and rare decay constraints \cite{Aaij:2013aka,Chatrchyan:2013bka,CMSandLHCbCollaborations:2013pla,bsmumuextra,btosgammalimits,bsgammaextra} almost rule out the MSSM. It is tempting therefore to explore well motivated extensions of the MSSM, such as NMSSM \cite{A_Djouadi} and BLSSM \cite{Khalil:2007dr}, which may alleviate the little hierarchy problem
of the MSSM through additional contributions to Higgs mass \cite{A_Djouadi,Elsayed:2011de} and also provide new DM candidates \cite{Cerdeno:2004xw} that may account for the relic density with no conflict with other phenomenological constraints.

In this article we analyze the constraints imposed by the Higgs mass limit and the gluino lower bound, which are the most stringent collider constraints, on the MSSM parameter space. In particular, these constraints imply that the gaugino mass, $m_{1/2}$, resides within the mass range:  $620~{\rm GeV} \lsim m_{1/2} \lsim 2000$~GeV. While the other parameters are much less constrained.  We study the effect of the measured DM relic density on the MSSM allowed parameter space. We emphasized that in this case all parameter space is ruled out except few points around $\tan \beta \sim 50$,  $m_0 \sim 1$~TeV and $m_{1/2} \sim 1.5$~TeV. We also investigate the direct detection rate of the LSP at these allowed points in light of the latest LUX result. Finally we show that if one assumes non-standard scenario of cosmology with low reheating temperature, where the LSP  may reach equilibrium before the reheating time, then the relic abundance constraints on $(m_0, m_{1/2})$ can be significantly relaxed. 

The paper is organized as follows. In section 2 we briefly introduce the MSSM and study the constraints on the $(m_0,m_{1/2})$ plane from Higgs and gluino mass experimental 
limits. In section 3 we study the thermal relic abundance of the LSP in the allowed region of parameter space. We show that the combined LHC and relic abundance constraints rule out most of the parameter space except the case of very large $\tan \beta$. We also provide the expected rate of direct LSP detection at these points with large $\tan \beta$ and TeV masses. Section 4 is devoted for non-thermal scenario of DM and how it can relax the constraints imposed on MSSM parameter space. Finally we give our conclusions in section 5.   

\section{MSSM After the LHC Run-I }

The particle content of  the MSSM is three generations of (chiral) quark and lepton superfields, the (vector)
superfields necessary to gauge the $SU(3)_C\times SU(2)_L\times U(1)_Y$ gauge of the
SM, and two (chiral) $SU(2)$ doublet Higgs superfields. The introduction of a second
Higgs doublet is necessary in order to cancel the anomalies produced by the fermionic
members of the first Higgs superfield, and also to give masses to both up and down type  
quarks. The interactions between Higgs and matter superfields are described by the 
superpotential
\begin{equation}
W = h_U Q_L U^c_L H_2 + h_D Q_L D_L^c H_1 + h_L L_L E_L^c H_1 +
\mu H_1 H_2. 
\label{superpot}
\end{equation}
Here $Q_L$ contains $SU(2)$ (s)quark doublets and $U_L^c$, $D_L^c$ are the corresponding singlets, (s)lepton doublets and singlets reside
in $L_L$ and $E_L^c$ respectively. While $H_1$ and $H_2$ denote
Higgs superfields with hypercharge $Y=\mp \frac{1}{2}$. 
Further, due to the fact that Higgs and lepton doublet superfields have
the same $SU(3)\times SU(2)_L\times U(1)_Y$ quantum numbers, we have
additional terms that can be written as 
\begin{equation}
W'= \lambda_{ijk}L_i L_j E^c_k + \lambda'_{ijk} L_i Q_j D^c_k
+\lambda''_{ijk}D^c_i D^c_j U^c_k + \mu_i L_i H_2.
\label{unwanted}
\end{equation}
These terms violate baryon and lepton number explicitly and lead to proton
decay at unacceptable rates. To forbid these terms a new symmetry, called $R$-parity,  
is introduced, which is defined as $R_P= (-1)^{3B+L+2S}$, where $B$ and $L$ are baryon 
and lepton number and $S$ is the spin.  There are two remarkable phenomenological implications 
of the presence of $R$-parity: $i)$ SUSY particles are produced or destroyed only in pair.
$ii)$ The LSP is absolutely stable and, hence, it might
constitute a possible candidate for DM.

\begin{figure}[t]
\begin{center}
\includegraphics[width=8cm,height=6cm]{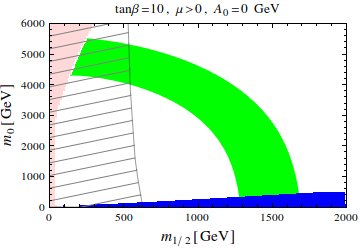}~~~~~\includegraphics[width=8cm,height=6cm]{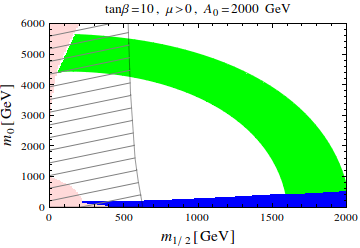}\\[0.5cm]
\includegraphics[width=8cm,height=6cm]{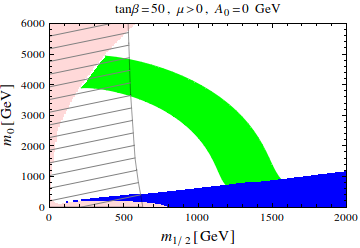}~~~~~\includegraphics[width=8cm,height=6cm]{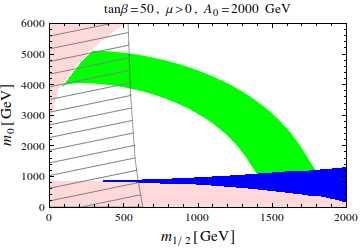}
\caption{MSSM parameter space for $\tan \beta=10$ (top panel) and $50$ (bottom panel) with $A_0=0$ and $2$~TeV. The green region indicates for $ 124 \lsim m_h \lsim 126$~GeV. The blue region is excluded because the lightest neutralino is not the LSP. The pink region is excluded due to absence of radiative electroweak symmetry breaking ($\mu^2$ becomes negative). The gray shadow lines denote the excluded area because of $m_{\tilde{g}} <1.4$~TeV.} 
\label{Fig1}
\end{center}
\end{figure}

In MSSM, a certain universality of soft SUSY breaking terms at grand unification scale
$M_X = 3 \times 10^{16}$~GeV is assumed. These terms are defined as $m_0$, the universal scalar soft mass,
$m_{1/2}$, the universal gaugino mass, $A_0$, the universal trilinear coupling, $B$, the bilinear coupling (the soft mixing 
between the Higgs scalars). In order to discuss the physical implication of soft SUSY breaking at low energy,
 we need to renormalize these parameters from $M_X$ down to electroweak scale. In addition the
 MSSM contains another two free SUSY parameters: $\mu$ and $\tan \beta =\langle H_2 \rangle/\langle H_1 \rangle$.  
 Two of these free parameters, $\mu$ and $B$, can be determined by the electroweak breaking conditions: 
\bea
\mu^2 &=& \frac{m_{H_1}^2 -m_{H_2}^2 \tan^2 \beta}{\tan^2\beta -1} -M_Z^2/2,
\label{minimize1}\\
 \sin 2\beta &=& \frac{-2 m_3^2}{m_1^2 + m_2^2}.
\label{minimize2}
\eea
Thus, the MSSM has only four independent free parameters: $m_0, m_{1/2}, A_0, \tan \beta$, besides to the sign of $\mu$, that 
determine the whole spectrum.

In the MSSM, the mass of the lightest Higgs state can be approximated, at the one-loop level, as \cite{Ellis} %
\be%
m_h^2 \leq M_Z^2 + \frac{3 g^2}{16\pi^2 M_W^2}
\frac{m_t^4}{\sin^2\beta} \log\left(\frac{m_{\tilde{t}_1}^2
m_{\tilde{t}_2}^2}{m_t^4}\right).%
\label{deltat}
\ee%
Therefore, if one assumes that the
stop masses are of order TeV, then the one-loop effect leads to a
correction of order ${\cal O}(100)$~GeV, which implies
that %
\be%
m_h^{\rm MSSM} \lsim \sqrt{(90~{\rm GeV})^2 + (100~{\rm GeV})^2} \simeq 135~{\rm GeV}.%
\ee%
The two-loop corrections reduce this upper bound by a few GeVs  \cite{Heinemeyer:1999be}. 
Hence, the MSSM predicts the following  upper bound for the Higgs mass: $m_h \lsim 130$~GeV, which was consistent 
with the measured value of Higgs mass (of order 125~GeV) at the LHC \cite{Aad:2012tfa}. 

In Fig.~\ref{Fig1} we display the contour plot of the SM-like Higgs boson: $m_h \in [124, 126]$~GeV in $(m_0, m_{1/2})$ plane for different values of $A_0$ and $\tan \beta$.
It is remarkable that the smaller $A_0$ is, the smaller $m_{1/2}$ is needed to satisfy this value of Higgs mass. It is also clear that the scalar mass $m_0$ remains essentially unconstrained by Higgs mass limit. It can vary from few hundred GeVs to few TeVs.  Such large values of $m_{1/2}$ seem to imply a quite heavy SUSY spectrum, much heavier that the lower bound imposed by direct searches at the LHC experiments in centre of mass energies $\sqrt s =7,8$~TeV and total integrated luminosity of order $20~{\rm fb}^{-1}$. Furthermore, the LHC lower limit on the gluino mass: $m_{\tilde{g}} \gsim 1.4$~TeV \cite{gluino search}, excluded the values of $m_{1/2} < 620$~GeV that was allowed by Higgs mass constraints for $m_0 > 4 $~TeV. Furthermore, this region is shown with dashed lines in Fig.~\ref{Fig1}.

\section{Dark Matter Constraints on MSSM parameter space}

\subsection{The LSP as dark matter candidate}

The neutralinos $\chi_i $ (i=1,2,3,4) are
the physical (mass) superpositions of two fermionic partners of the two
neutral gauge bosons, called gaugino $\tilde{B}^0$ (bino) and
$\tilde{W}^0_3$ (wino), and of the two neutral Higgs bosons, called
Higgsinos $\tilde{H}_1^0$ and $\tilde{H}_2^0 $. The neutralino mass
matrix is given by~\cite{neutralino}

{\small 
\begin{equation}
M_N = \left(\begin{array}{cccc}
M_1&0&-M_Z\cos\beta\sin\theta_W&M_Z\sin\beta\sin\theta_W\\
0&M_2&M_Z\cos\beta\cos\theta_W&-M_Z\sin\beta\cos\theta_W\\
-M_Z\cos\beta\sin\theta_W&M_Z\cos\beta\cos\theta_W&0&-\mu\\ 
M_Z\sin\beta\sin\theta_W&-M_Z\sin\beta\cos\theta_W&-\mu&0
\end{array}\right),
\label{neutralino}
\end{equation}
}
where $M_1$ and $M_2$ are related due to the universality of the gaugino
masses at the grand unification scale,
$ M_1= \frac{3 g_1^2}{5 g_2^2} \ M_2$, where $g_1$, $g_2$ are the gauge couplings of $U(1)_Y$ and $SU(2)_L$ respectively. This Hermitian matrix is diagonlaized by a unitary transformation of the
neutralino fields, $M_N^{diag}= N^{\dag} M_N N$.
The lightest eigenvalue of this matrix and the corresponding eigenstate
say $\chi$ has good chance of being the LSP. The lightest neutralino will be a linear
combination of the original fields:
\begin{equation}
\chi = N_{11}\tilde{B}^0+ N_{12}\tilde{W}^0+
N_{13}\tilde{H}_1^0 + N_{14}\tilde{H}_2^0.   
\end{equation}
The phenomenology and cosmology of the neutralino are governed primarily
by its mass and composition. A useful parameter for describing the
neutralino composition is the gaugino ``purity" function $ f_g=\vert
N_{11}\vert^2 +\vert N_{12} \vert^2$ ~\cite{neutralino}. If $f_g > 0.5$, then
the neutralino is primarily gaugino and if $f_g < 0.5$, then the
neutralino is primarily Higgsino. Actually if $\vert \mu \vert > \vert
M_2 \vert \geq M_Z $, the two lightest neutralino states will be
determined by the gaugino components, similarly, the light chargino will 
be mostly a charged wino. While if $\vert \mu \vert < \vert M_2 \vert$,
the two lighter neutralinos and the lighter chargino are all mostly
Higgsinos, with mass close to $\vert \mu \vert$. Finally if $\vert \mu
\vert \simeq \vert M_2\vert $, the states will be strongly mixed.

\begin{figure}[t]
\includegraphics[width=8cm,height=7cm]{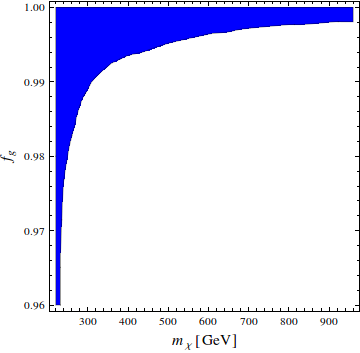}
\caption{The mass of lightest neutralino versus the purity function in the region of parameter space allowed by 
gluino and Higgs mass limits.} \label{fg}
\end{figure}

Here, two remarks are in order: $i)$ The above mentioned constraints in $m_{1/2}$ from Higgs mass limit and gluino 
mass lower bound imply that $m_{\chi} \gsim 240$~GeV, which is larger than the limits obtained from direct searches at the LHC. 
Moreover, an upper bound of order one TeV is also obtained (from Higgs mass constraint). $ii)$ In this region of allowed parameter space, 
the LSP is essentially pure bino, as shown in Fig.~\ref{fg}. This can be easily understood from the fact that $\mu$-parameter, determined by 
the radiative electroweak breaking condition, Eq.~(\ref{minimize1}), is typically of of order $m_0$ and hence it is much heavies than the gaugino mass $M_{1}$. 

\subsection{Relic denisty}

As advocated in the previous section, the LSP in MSSM, the lightest neutralino  $\chi$, 
is a perfect candidate for DM.  Here, we assume that  $\chi$ was in thermal equilibrium with the SM particles 
in the early universe and decoupled when it was non-relativistic.
Once the $\chi$ annihilation rate $\Gamma_{\chi} =\langle
\sigma^{ann}_{\chi}\ v \rangle n_{\chi}$ dropped below the
expansion rate of the universe, $\Gamma_{\chi} \leq H$, the LSP particles
stop to annihilate, fall out of equilibrium and their relic
density remains intact till now. The above $\langle
\sigma^{ann}_{\chi}\ v \rangle$ refers to thermally averaged total
cross section for annihilation of $\chi \chi$ into lighter
particles times the relative velocity, $v$.

The relic density is then determined by the Boltzmann equation for
the LSP number density $(n_{\chi})$ and the law of entropy
conservation: 
\bea 
\frac{d n_{\chi}}{dt} &=& -3 H n_{\chi}  -
\langle \sigma^{ann}_{\chi}\ v \rangle \left[(n_{\chi})^2 -
(n^{eq}_{\chi})^2\right],\\
\frac{d s}{dt} &=& - 3 H s, \label{S(T)}
\eea 
where $n_{\chi}^{eq}$ is the LSP equilibrium number density which, as function of
temperature $T$, is given by $n^{eq}_{\chi} = g_{\chi} (m_{\chi}
T/2\pi)^{3/2} e^{-m_{\chi}/T}$. Here $m_{\chi}$ and $g_{\chi}$ are
the mass and the number of degrees of freedom of the LSP
respectively. Finally, $s$ is the entropy density. In the standard
cosmology, the Hubble parameter $H$ is given by $H(T)=2\pi\sqrt{\frac{\pi g_*}{45}} \frac{T^2}{M_{Pl}}$, where $M_{Pl}=1.22\times 10^{19}$~GeV and $g_*$ is the number of relativistic degrees of freedom. Let us introduce the variable $x =m_{\chi}/T$ and define $Y
=n_{\chi}/s$ with $Y_{eq} =n^{eq}_{\chi}/s$. In this case, the
Boltzmann equation is given by  
\be 
\frac{dY}{dx} = \frac{1}{3H}
\frac{d s}{dx} \langle \sigma^{ann}_{\chi}\ v \rangle \left(Y^2 -
Y^2_{eq}\right).
\ee 
In radiation domination era, the entropy, as
function of the temperature, is given by 
\be 
s(x) = \frac{2\pi^2}{45} g_{\ast_s}(x)~ m_{\chi}^3~ x^{-3},
\ee
which is deduced from the fact that $s=(\rho+ p)/T$ and $g_{\ast_s}$ is the
effective degrees of freedom for the entropy density. Therefore
one finds 
\be
 \frac{d s}{d x} = - \frac{3 s}{x}.
 \ee 
Thus, the following expression for the Boltzmann equation
for the LSP number density is obtained 
\be 
\frac{dY}{dx} = -\sqrt{\frac{\pi g_*}{45}} M_{Pl}~
m_{\chi} \frac {\langle \sigma^{ann}_{\chi}\ v \rangle }
{x^2}\left(Y^2 - Y^2_{eq}\right).\label{Boltzmann1}\ee

If one considers the s-wave and p-wave annihilation processes only, the thermal average $\langle\sigma^{ann}_{\chi}\ v \rangle$ then shows as 
\be
\langle\sigma^{ann}_{\chi} \ v \rangle=a_{\chi}+\frac{6b_{\chi}}{x},
\ee
where $a_{\chi}$ and $b_{\chi}$ are the s-wave and p-wave contributions of annihilation processes, respectively. The relic density of the DM candidate is given by
\begin{equation}
\Omega h^2=\frac{m_{\chi} s_0 Y_{\chi}(\infty)}{\rho_c/h^2},
\end{equation}
where $s_0=2282.15\times 10^{-41}$~GeV$^3$, $\rho_c=8.0992\,h^2\times 10^{-47}$~GeV$^4$, and by solving the Boltzmann equation, one can find $Y_{\chi}(\infty)$ as follows \cite{Kolb}
\begin{equation}
Y_{\chi}(\infty)= \frac{1}{\lambda_{\chi}}\left(\frac{a_{\chi}}{x(T_f)}+\frac{3b_{\chi}}{x^2(T_f)}\right)^{-1},
\end{equation}
where $T_f$ is the freeze-out temperature, $\lambda_\chi=s(m_\chi)/H(m_\chi)$ and $x(T_f)$ is given by  
\begin{equation}
x(T_f)=\ln \left[\frac{\alpha_{\chi} \lambda_{\chi} c(c+2)}{\sqrt{x(T_f)}}\left( a_{\chi} +\frac{6b_{\chi}}{x(T_f)}\right)\right],
\end{equation}
where $\alpha_{\chi}=\frac{45}{2\pi^4}\sqrt{\frac{\pi}{8}}\frac{g_{\chi}}{g_{\ast_s}(T_f)}$, the value $c=\frac{1}{2}$ results in a typical accuracy of about $5-10 \% $ more than sufficient for our purposes here.

\begin{figure}[t]
\includegraphics[scale=0.6]{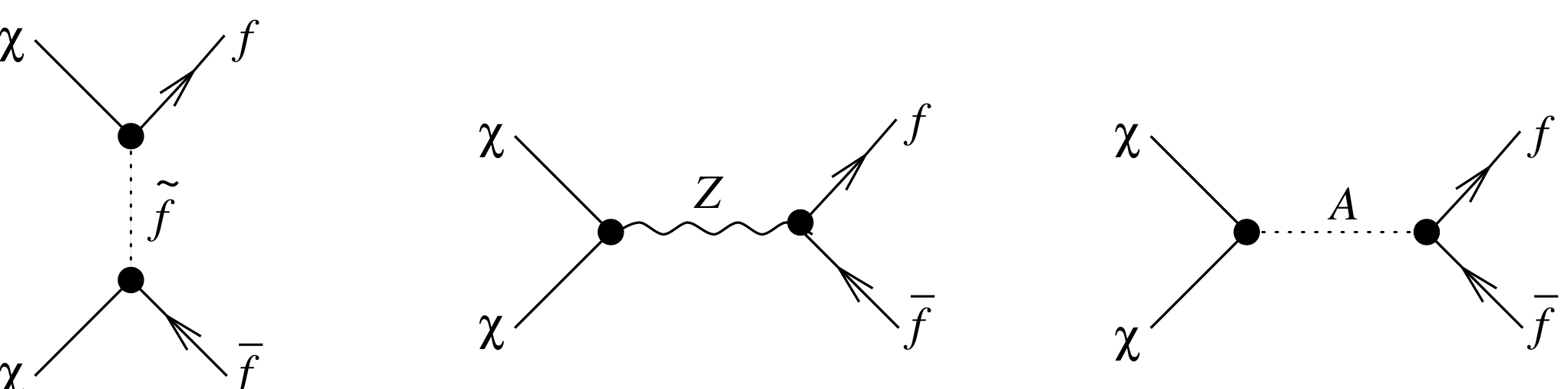}
\caption{Feynman diagrams contributing to early-universe
neutralino ($\tilde{\chi}^0_1$) annihilation into fermions through
sfermions, $Z$-gauge boson, and Higgs.} \label{annihilation}
\end{figure}

The lightest neutralino may annihilate into fermion-antifermion ($f\bar{f}$), $W^+ W^-$, $Z
Z$, $W^+ H^-$, $Z A$, $Z H$, $Z h$, $H^+ H^-$ and
all other contribution of neutral Higgs. For a bino-like LSP, \ie\,
$N_{11} \simeq 1$ and $N_{1i}\simeq 0$, $i=2,3,4$, one finds that
the relevant annihilation channels are the fermion-antifermion
ones, as shown in Fig.~\ref{annihilation}, and all other channels
are instead suppressed. Also, the annihilation process mediated by $Z$
gauge boson is suppressed due to the small $Z \chi \chi$ coupling
$\propto N_{13}^2 - N_{14}^2$, except at the resonance when
$m_\chi \sim m_Z/2$, which is no longer possible due to the above mentioned constraints. Furthermore, one finds that the annihilation
is predominantly into leptons through the exchanges of the three
slepton families $(\tilde{l}_L, \tilde{l}_R)$, with $l=e,\mu,\tau$. The squarks
exchanges are suppressed due to their large masses.

\begin{figure}[t]
\begin{center}
\includegraphics[width=8cm,height=6cm]{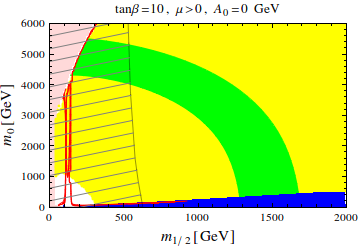}~~~~~\includegraphics[width=8cm,height=6cm]{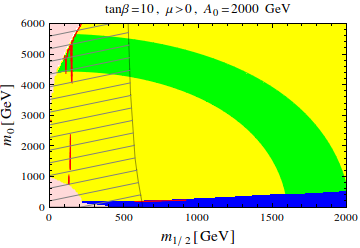}\\[0.5cm]
\includegraphics[width=8cm,height=6cm]{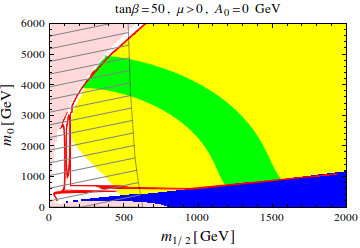}~~~~~\includegraphics[width=8cm,height=6cm]{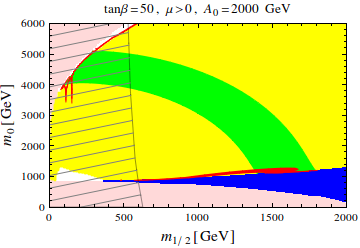}
\caption{LSP relic abundance constraints (red regions) on $(m_0-m_{1/2})$ plane for $\tan \beta$ and $A_0$ as in Fig.~\ref{Fig1}. The LUX result is satisfied by the yellow region. The other color codes are as in Fig.~\ref{Fig1}. }
\label{Fig2}
\end{center}
\end{figure}

\begin{figure}[t]
\begin{center}
\includegraphics[width=8cm,height=6cm]{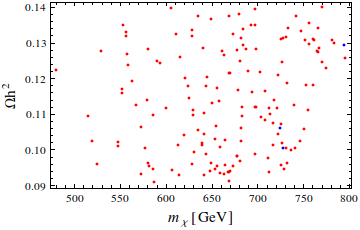}
\caption{The relic abundance versus the mass of the LSP for different values of $\tan{\beta}$. Red points indicate for $40\leq \tan{\beta}\leq 50$ and blue points for $30\leq \tan{\beta}<40$. All points satisfy the above mentioned constraints.} \label{Fig3}
\end{center}
\end{figure}

In Fig.~\ref{Fig2} we display the constraint from the observed limits of $\Omega h^2$ on the plane $(m_0-m_{1/2})$ for $A_0=0, 2000$~GeV, $\tan \beta= 10, 50$ and $\mu >0$. Here we used micrOMEGAs \cite{Belanger:2013oya} to compute the complete relic abundance of the lightest neutralino, taking into account the possibility of having co-annihilation with the next-to-lightest supersymmetric particle, which is typically the lightest stau. In this figure the red regions correspond to a relic abundance within the measured limits \cite{Ade:2013zuv}:
\be
0.09<\Omega h^2<0.14
\ee
It is noticeable that with low $\tan\beta~(\sim 10)$, this region corresponds to light $m_{1/2}$ $(< 500$~GeV$)$, where a significant co-annihilation between the LSP and stau took place. However, this possibility is now excluded by the Higgs and gluino mass constraints \cite{Chakraborti:2014fha}. At large $\tan\beta$, another region is allowed due to a possible resonance due to s-channel annihilation of the DM pair into fermion-antifermion via the pseudoscalar Higgs boson $A$ at $M_A\simeq 2 m_{\chi}$ \cite{Kowalska:2013hha}. For $A_0=0$, a very small part of this region is allowed by the Higgs mass constraint, while for large $A_0~(\sim 2$~TeV$)$ a slight enhancement of this part can be achieved. In Fig.~\ref{Fig3}, we zoom in on this region to show the explicit dependence of the relic abundance on the LSP mass and large values of $\tan{\beta}$. As can be seen from this figure,  there is no point can satisfy the relic abundance stringent constraints with $\tan{\beta}<30$.

\subsection{Direct Detection}

Perhaps the most natural way of searching for the neutralino DM is provided
by direct experiments, where the effects induced in appropriate detectors by neutrali-nonucleus
elastic scattering may be measured. The elastic-scattering cross section of
the LSP with a given nucleus has two
contributions: spin-dependent contribution arising from $Z$ and
$\tilde{q}$ exchange diagrams, and spin-independent (scalar)
contribution due to the Higgs and squark exchange diagrams, which is typically suppressed. The
effective scalar interaction of neutralino with a quark is given by
\be%
{\cal L} = a_q \bar{\chi} \chi \, \bar{q} q, %
\ee%
where $a_q$ is the neutralino-quark effective coupling. The scalar cross section of the neutralino scattering with target
nucleus is given by \cite{Jungman:1995df}%
\be%
\sigma_{\rm SI} = \frac{4 m_r^2}{\pi} \left(Z f_p + (A-Z) f_n \right)^2,%
\ee%
where $Z$ and $A-Z$ are the usual atomic numbers, $m_r$ is the
reduced mass of the nucleon and $f_p$, $f_n$ are the neutralino coupling
to protons and neutrons respectively.

\begin{figure}[t]
\begin{center}
\includegraphics[width=8cm,height=6cm]{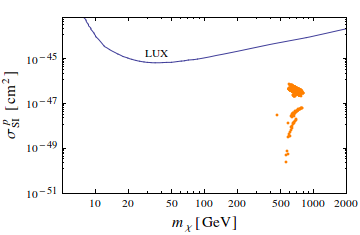}
\caption{Spin-independent scattering cross section of the LSP with a proton versus the mass of the LSP within the region allowed by all constraints (from the LHC and relic abundance). } \label{Fig5}
\end{center}
\end{figure}

In Fig.~\ref{Fig5} we display the MSSM prediction for spin-independent scattering cross section of the LSP with a proton after imposing the LHC and relic abundance constraints. It is clear that our results for $\sigma_{SI}^p$ are less than the recent LUX bound (blue curve) by at least two order of magnitudes. This would explain the negative results of direct searches so far.

\section{Non-thermal Dark Matter and MSSM parameter space }

\begin{figure}[t]
\begin{center}
\includegraphics[width=8cm,height=6cm]{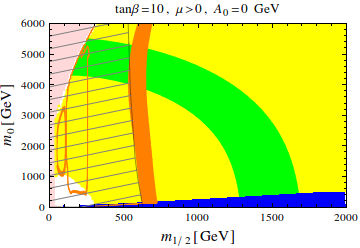}~~~~~\includegraphics[width=8cm,height=6cm]{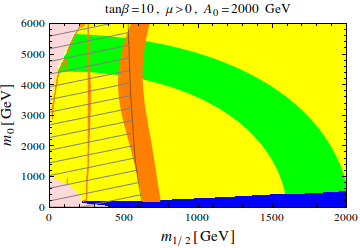}\\[0.5cm]
\includegraphics[width=8cm,height=6cm]{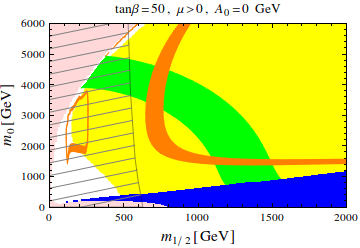}~~~~~\includegraphics[width=8cm,height=6cm]{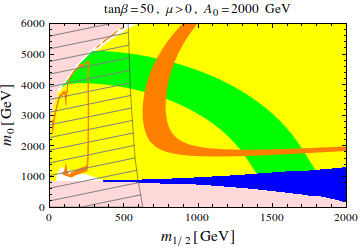}
\caption{LSP non-thermal relic abundance constraints (red regions) on $(m_0-m_{1/2})$ plane for $\tan \beta$ and $A_0$ as in Fig.~\ref{Fig1}. The color codes are as in Fig.~\ref{Fig1}.}
\label{non-thermal}
\end{center}
\end{figure}

In the previous section, we assumed standard cosmology scenario where the reheating temperature
$T_{RH}$ is very large, namely $T_{RH}>>T_f\sim 10$~GeV. However, the only constraint on the reheating temperature, 
which could be associated with decay of any scalar field, $\phi$, not only the inflaton field, is $T_{RH}\gsim 1$~MeV in order not to spoil the successful  
predictions of big bang nucleosynthesis. 

A detailed analysis of the relic density with a low reheating
temperature has been carried out in Ref.~\cite{Giudice:2000ex}.  
It was emphasized that for a large annihilation
cross section, $\langle \sigma^{ann}\ v \rangle  \gsim 10^{-14}$~GeV$^{-2}$ so that the neutralino reaches equilibrium before reheating, and if there is a large number of neutralinos
produced by the scalar field $\phi$ decay, then the relic density is estimated as \cite{Moroi:1999zb}
\begin{eqnarray}
\Omega h^2 & =& \frac{3\ m_{\chi}\ \Gamma_{\phi}}
{2\ (2 \pi^2/45)\ g_*\ T_{RH}^3\  \langle 
\sigma_\chi^{ann}\ v  \rangle}\ 
\frac{h^2}{ \rho_c /s_0}.
\label{nlsp}
\end{eqnarray}
Here the reheating temperature is defined as~\cite{Kolb} 
\be 
T_{RH}= \left(\frac{90}{\pi^2 g_*(T_{RH})}\right)^{1/4} 
(\Gamma_\phi M_{Pl})^{1/2}.
\label{trh}
\ee 
where the decay width $\Gamma_\phi$ is given by
\be
\Gamma_{\phi} = \frac{1}{2 \pi} \frac{m_{\phi}^3}{\Lambda^2}.
\label{gphi}
\ee
The scale $\Lambda$ is the effective suppression scale, which is of order the grand unification scale $M_X$. Therefore, for scalar field with mass $m_\phi \simeq 10^7$~GeV one finds $\Gamma_\phi \simeq 10^{-11}$~GeV and in our calculations, we have used $g_*=10.75$ due to the consideration of a low reheating temperature scenario.

In Fig.~\ref{non-thermal} we show the constraints imposed on the MSSM $(m_0-m_{1/2})$ plane in case of non-thermal relic abundance of the LSP for $\tan \beta=10,50$ and $A_0=0,2$~TeV. In this plot, we also imposed the LHC constraints, namely the Higgs mass limit and the gluino mass lower bound, similar to the case of thermal scenario. It is clear from this figure that the stringent constraints imposed of the MSSM parameter space by thermal relic abundance are now relaxed and  now low $\tan \beta ~(\sim 10)$ is allowed but with very heavy $m_0$ ($\sim  {\cal O}(4)$~TeV$)$ and $m_{1/2}\sim 600$~GeV. In addition, the following two regions are now allowed with large $\tan \beta ~(\sim 50)$:  
$(i)~ m_0 \sim m_{1/2} \sim {\cal O}(1)$~TeV; $(ii)~ m_0 \simeq {\cal O}(4)$~TeV and $m_{1/2} \simeq 700$~GeV. The SUSY spectrum associated with these regions of parameters space could be striking signature for non-thermal scenario at the LHC.

\section{Conclusion}

We have studied the constraints imposed on the MSSM parameter space by the Higgs mass limit and the gluino lower bound, which are the most stringent collider constraints obtained from the LHC run-I at energy 8~TeV.  We showed that $m_{1/2}$ resides within the mass range: $620~{\rm GeV} \lsim m_{1/2} \lsim 2000$~GeV, while the other parameters ($m_0$, $A_0$, $\tan\beta)$ are much less constrained.  We also studied the effect of the measured DM relic density on the MSSM allowed parameter space. It turns out that most of the MSSM parameter space is ruled out except few points around $\tan \beta \sim 50$, $m_0 \sim 1$~TeV and $m_{1/2} \sim 1.5$~TeV. We calculated the spin-independent scattering cross section of the LSP with a proton in this allowed region. We showed that our prediction for $\sigma_{SI}^p$ is less than the recent LUX bound by at least two order of magnitudes.  We have also analyzed the non-thermal DM scenario  for the LSP. We showed that the constraints imposed on the MSSM parameter space is relaxed and low $\tan \beta$ is now allowed with $m_0\simeq {\cal O}(4)$~TeV and $m_{1/2}\simeq 600$~GeV. Also  two allowed regions are now associated with large $\tan \beta ~(\sim 50)$, namely: $m_0\sim m_{1/2} \simeq {\cal O}(1)$~TeV or $m_0 \simeq {\cal O}(4)$~TeV and $m_{1/2} \simeq 700$~GeV.

\section*{Acknowledgements}

This work was partially supported by the STDF project 13858, the ICTP grant AC-80 and the  European Union FP7  ITN INVISIBLES (Marie Curie Actions, PITN-GA-2011-289442).

\end{document}